\documentclass[journal=jctcce,manuscript=article]{achemso}

\usepackage[version=3]{mhchem} 
\usepackage{amsmath}
\usepackage{amsfonts}
\usepackage{amssymb}
\usepackage{mathtools}
\usepackage{color}

\newcommand{\tqq}{T_{q_i,q_{i+1}}}
\newcommand{\tqqrev}{T_{q_{i+1},q_i}}

\newcommand{\msm}{Chodera2007,Noe2007, Bowman2009,Noe2009,Bowman2010,Prinz2011,BowmanMSMBook}

\DeclareMathOperator*{\argmin}{arg\,min}
\DeclareMathOperator*{\argmax}{arg\,max}
\author{Ernesto Su\'{a}rez}
\email{ernesto.suarezalvarez@nih.gov}
\affiliation{National Cancer Institute, Frederick, Maryland}
\author{Daniel M. Zuckerman}
\email{zuckermd@ohsu.edu}
\affiliation{OHSU Center for Spatial Systems Biomedicine and Department of Biomedical Engineering, Oregon Health and Science University, Portland, Oregon}

\title{Pathway Histogram Analysis of Trajectories: A general strategy for quantification of molecular mechanisms}

\abbreviations{MSM, Markov state model; NM, non-Markovian; RMSD, root mean-squared deviation}

\begin{document}

\begin{abstract}
A key overall goal of biomolecular simulations is the characterization of ``mechanism'' -- the pathways through configuration space of processes such as conformational transitions and binding.  
Some amount of heterogeneity is intrinsic to the ensemble of pathways, in direct analogy to thermal configurational ensembles.
Quantification of that heterogeneity is essential to a complete understanding of mechanism.
We propose a general approach for characterizing path ensembles based on mapping individual trajectories into  pathway classes whose populations and uncertainties can be analyzed as an ordinary histogram, providing a quantitative ``fingerprint'' of mechanism.
In contrast to prior flux-based analyses used for discrete-state models, stochastic deviations from average behavior are explicitly included via direct classification of trajectories.
The histogram approach, furthermore, is applicable to analysis of continuous trajectories.
It enables straightforward comparison between ensembles produced by different methods or under different conditions. 
To implement the formulation, we develop approaches for classifying trajectories, including a clustering-based approach suitable for both continuous-space (e.g., molecular dynamics) or discrete-state (e.g., Markov state model) trajectories, as well as a ``fundamental sequence'' approach tailored for discrete-state trajectories but also applicable to continuous trajectories through a mapping process.  
We apply the pathway histogram analysis to a toy model and an extremely long atomistic molecular dynamics trajectory of protein folding.
\end{abstract}

\section{Introduction}

Protein conformational processes, which include folding and conformational transitions, are essential to innumerable cellular functions \cite{Alberts-2002}, and a key goal of computer simulations is to quantify such processes \cite{Zuckerman2011,Chong2017,Chandler-2002}.  Characterizing the ``mechanisms'' (pathways) of such processes, the series of conformational states through which a system transits during the process of interest, is at the heart of the biomolecular simulation enterprise.

Because straight-ahead molecular dynamics (MD) simulations generally extend to timescales that are only a fraction of what is required to exhibit biologically significant conformational processes \cite{Karplus2002,Zuckerman2011}, specialized computational methods have been developed to infer long-time behavior from affordable simulations.  
A number of ``path sampling'' techniques have been developed in recent decades which generally attempt to construct continuous or nearly continuous trajectory ensembles based on principles of non-equilibrium statistical mechanics \cite{Chong2017, Bello-Rivas2015,warmflash2007umbrella,allen2006simulating}.
Further, path optimization approaches estimate a single dominant pathway \cite{Maragliano2006,Henkelman2000}, and based on a set of short trajectories, Markov state models (MSMs) use statistical inference to ``stitch together'' mechanistic pathways in a reduced configurational description consisting of discrete states \cite{\msm}.

The analysis and interpretation of pathways constitutes another challenge, since their intrinsic structure is often hidden behind the stochastic nature of the system and the limited sampling. 
Our overall strategy is to classify pathways in a well-defined way that enables (i) statistical comparison to another path ensemble, perhaps of unbiased reference data, (ii) ready analysis of the dominant transition mechanisms and the degree of heterogeneity in a path ensemble, and (iii) applicability to continuous or discrete trajectories. 
We always assume that the macrostates of interest, A and B (e.g., folded and unfolded) have already been defined but the discussion applies to arbitrary macrostates. To classify the pathway ensembles from any approach (e.g., MSM or MD trajectories), we require a scheme that maps trajectories to pathway classes in a physically reasonable way.

For analysis of MSMs, a flux-based pathway classification strategy \cite{Noe2009, Hummer2009} has been widely used.
The approach uses an intuitive mapping of flux pathways that allows the fractional flow through each path to be computed.
Fluxes, however, represent \emph{average} behavior and simply following flux lines excludes stochastic but expected switching between (non-crossing) flux lines, as described below.
Of equal importance, flux lines cannot be used to \emph{classify} actual trajectories because of the stochastic behavior just noted.
In general, even a sequence of MSM states generated probabilistically from the MSM transition matrix could not be uniquely assigned to a single flux path.
Of equal importance here, MD trajectories, even if mapped to a discrete MSM space, also could not be classified by flux-based paths.

\begin{figure}[h] 
\begin{center}
  \includegraphics[width=1.0\linewidth]{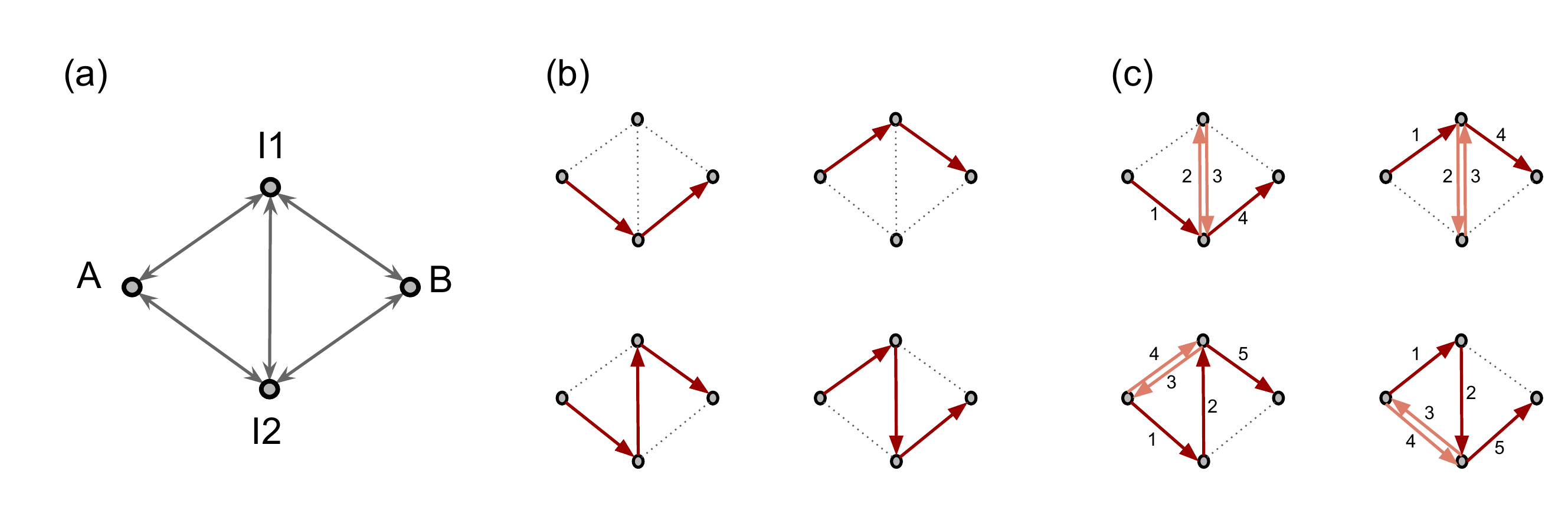}
  \caption{Simple model kinetic model. (a) Kinetic model with four states, the arrows are drawn between states kinetically connected. (b) Possible mechanisms from $A$ to $B$. (c) Examples of additional paths with ``unproductive'' steps.}
  \label{fig:ToyModel}
\end{center}
\end{figure}

A very simple model can illustrate the complexity of the path ensemble and suggest strategies to simplify its analysis. Suppose we have a very coarse partition of the configurational space with only four states; a initial state $A$, a final state $B$, and two intermediate states $I1$ and $I2$ that connect $A$ and $B$ as shown in Fig.~\ref{fig:ToyModel}a. Following our ``chemical intuition'' we can identify four different mechanisms to go from $A$ to the final state $B$ (Fig.~\ref{fig:ToyModel}b). However, the path ensemble is much more complex, having an unlimited number of paths (examples in Fig.~\ref{fig:ToyModel}c) . We could also identify an infinite family of paths, for instance, sequences following the pattern $A \rightarrow (I1 \rightarrow I2 \rightarrow \mathrm{A} \rightarrow I1)_n \rightarrow B$, for any positive integer $n$.

In order to propose a mechanism we need to reduce the dimensionality of the problem.  A simplified description could be given by how the reactive flux from $A$ to $B$ is distributed over the network of ``edges'' that connect the states \cite{Noe2009, Hummer2009}. However, a reactive flux description averages out potentially important elements of the path ensemble.  In our previous example, for instance, each edge in Fig~\ref{fig:ToyModel}a would be unidirectional in a flux picture, and the I1-I2 edge would have zero flux in a symmetric system, thus excluding possible paths that move transversely or backwards and failing to separate distinct pathways that may share common segments. 
We want a description of the path ensemble without discarding possible paths, including sub-optimal ``orthogonal-space'' transitions that occur in realistic trajectories.

\begin{figure}[h] 
\begin{center}
  \includegraphics[width=0.5\linewidth]{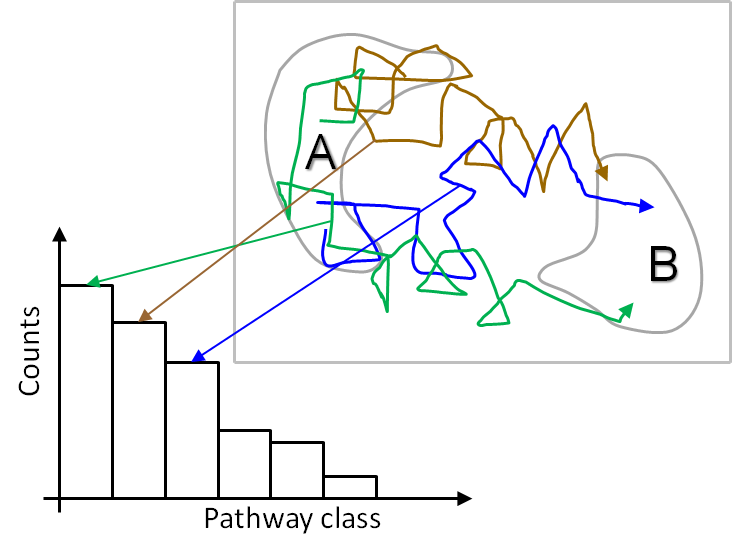}
  \caption{Pathway histogram analysis of trajectories.  Trajectories are mapped to distinct pathway classes and then a histogram is constructed.  The approach provides a quantitative ``fingerprint'' of mechanism and enables statistical comparison of trajectory ensembles.}
  \label{fig:histo}
\end{center}
\end{figure}

To better describe the mechanism and evaluate the performance of possible models, we propose to map the entire path ensemble to a limited number of classes: see Fig.\ \ref{fig:histo}. Each class would be a family of paths with a likelihood easy to quantify. The probability distribution (histogram) over the classes would act like the fingerprint of the path ensemble. Once the mapping function is defined, it would be very easy to compare the actual MD simulation with those generated from models and evaluate their performance; we just have to map them all with the same function. To summarize, the classification will allow us to compare complex ensembles via the comparison of simple histograms.

In the remainder of the paper, we demonstrate how the histogramming strategy can be applied to both discrete-state and continuous trajectories.
For discrete-state trajectories, such as from MSMs, histograms can be constructed via direct mapping from trajectory to pathway class or indirectly by clustering and then classifying.
We introduce a direct mapping which yields what we term the ``fundamental sequence'' --- roughly, the discrete-state trajectory with loops removed.
We further demonstrate that pathway classes can be generated from arbitrary clusterings of paths via a Voronoi decomposition -- where the initial clustering can be performed based on any algorithm and distance metric.
Continuous trajectories can be classified indirectly using the clustering-to-classes approaches or by mapping the trajectories to a discrete basis and then applying the fundamental-sequence -- or any other -- classification.



\section{Methods}\label{sec:methods}
Grouping together ``similar'' pathways is probably the simplest way to describe the mechanism of a biomolecular process. 
The goal is to expose the ``structure'' of the path ensemble through a regular classification procedure. In this section we will first show an automated classification scheme based on what we called the \emph{fundamental sequence} (FS) which is designed for discrete trajectories, e.g., the trajectories we would obtain from MSMs. Then, we will discuss how an initial clustering in path space can be used as starting point for classification of both discrete and continuous pathways.

\subsection{Classification based on direct mapping to pathway classes}
If every transition trajectory can be directly mapped to one of a discrete set of pathway classes, a path histogram is created. Such a histogram can be used for mechanistic analysis in its original form, or the class bins can be combined using a similarity metric for a more coarse grained description.

The classes themselves can be generated in different ways. 
They could be based on physical intuition -- for example, by considering the sequences of certain discrete events of interest which occur during the transitions.

Here we present an automated classification procedure for trajectories in discrete spaces based on the fundamental sequence (FS) idea.
Such discrete trajectories might result from mapping detailed configurations to a discrete set of configuration-space regions, or from some dynamics defined in the discrete space, such as a Markov state model.
Roughly speaking, the FS of a discrete path is what remains after eliminating the ``loops'' in the path. A loop is any sub-segment of the path where the initial and final states are the same (do not confuse with initial and final \emph{macro} states). In Fig.~\ref{fig:ToyModel}c, for instance, the segment of the path with arrows in light red are loops. Even self transitions --when the systems stays in the same state after $\tau$-- can be considered loops of size zero. If two paths share the same FS, then they will belong to the same class. There is not, however, a unique way to remove the loops and we need an unambiguous definition for FS.

Imaging drawing a graph $G(\mathbf{s})$ from the \emph{single} discrete trajectory/path $\mathbf{s} = (s_1, s_2, ...,s_m)$ from $A$ to $B$ ($s_1 \in A$, $s_m \in B$), where the nodes of the graph would be all the $N$ states. 
The graph edges are built following one single rule: if there is one or more \emph{direct} transitions from $m$ to $l$ in $\mathbf{s}$ (i.e., $\exists \, i: s_i=m\; \mathrm{and}\; s_{i+1}=l$ ), a \emph{directed} edge $m\rightarrow l$ is added to $G$. 
At the end, some edges will have a single direction ($m \rightarrow l$), others will be bidirectional ($m \rightleftarrows l$) and there will also be in general isolated nodes not visited by the particular path $\mathbf{s}$.
We assume all edges have known transition probabilities $T_{ml}$, which for molecular systems can be approximated using MD trajectories as is done in MSM construction \cite{\msm}.

The \emph{fundamental sequence} (FS) of a path $\mathbf{s}$ is the connected linear subgraph $\mathbf{q}$ of nodes in $G(\mathbf{s})$ that maximizes the product of the transition probabilities through all possible paths $\Gamma (G)$ consistent with $G(\mathbf{s})$.  That is

\begin{equation}\label{eq:FS}
\mathrm{FS}(\mathbf{s})= \argmax_{\mathbf{q} \in \Gamma(G)} \prod_{i=1}^{|\mathbf{q}|-1} \tqq.
\end{equation}
where $|\mathbf{q}|$ is the number of elements in the path $\mathbf{q} \in \Gamma(G(\mathbf{s}))$, and $|\mathbf{q}|-1$ is the number of transitions. 

An equivalent and more convenient formulation would be 

\begin{equation}\label{eq:FSlog}
\mathrm{FS}= \argmin_{\mathbf{q} \in \Gamma(G)} \left \{  \sum_{i=1}^{|\mathbf{q}|-1} -\log \tqq \right \},
\end{equation}
where have transformed the problem of maximizing the product of the transition probabilities to the classical problem of finding the shortest distance in a graph, in this case, using the pseudo distance 
$\delta(m,l) = -\log T_{ml}$. 
There are well known strategies to solve this kind of problem; here we use Dijkstra's algorithm \cite{Dijkstra1959}. 

A simple example can be seen in Fig.~\ref{fig:ToyModel}, where the FSs of the paths in Fig.~\ref{fig:ToyModel}c are the corresponding sequences in Fig.~\ref{fig:ToyModel}b. In general, any possible path from $A$ to $B$ in that simple model will correspond to one of the four \emph{fundamental sequences} drawn in Fig.~\ref{fig:ToyModel}b.

For the description of the mechanism, and in particular for equilibrium ensembles which exhibit mechanistic symmetry for forward and reverse directions \cite{bhatt2011beyond}, it is more natural to use a partition of path space which is symmetric by construction. That is, for a system in equilibrium, the distribution of the folding paths over the classes should be equal to the distribution of the unfolding paths on the same space. Instead of maximizing the likelihood of the sequence given the connectivity of the path, we will maximize the likelihood of the ``round trip''. The symmetrized fundamental sequence FS$^*$ is therefore defined as

\begin{equation}\label{eq:FSlogSymm2}
\mathrm{FS}^*= \argmin_{\mathbf{q} \in \Gamma(G)} \left \{  \sum_{i=1}^{|\mathbf{q}|-1} -\log ( \tqq \tqqrev ) \right \}.
\end{equation}

From now on, when me mention the fundamental sequence, we really mean this symmetrized version. 


Lastly we note that the prescriptions above for computing the FS are not themselves unique and could be modified to normalize for path length and/or employ a more explicit maximum flux definition.
Regardless of the definition, such FS classes provide a means for binning arbitrary stochastic discrete-state trajectories and creating a histogram.

\subsection{Classification based on clustering in path space - for discrete or continuous trajectories} \label{sec:classVoroGeneral}
To generate a path histogram, we require true \emph{classification} of path space into non-overlapping, space-tiling regions, which typically is not achieved using clustering methods \cite{barbakh2009review}.
However, it is straightforward to use arbitrary clusters and an arbitrary distance metric in path space to generate Voronoi-cell classification of the space.
Most simply, a representative member of each cluster can be taken as the reference structure for a Voronoi cell defined using the distance metric.
Although the distance metric used for clustering and Voronoi classification need not be the same, here we will assume they are identical.

Importantly, because clustering can be performed for either continuous or discrete trajectories, the corresponding Voronoi classification is equally general.
We will demonstrate clustering-based classification using discrete trajectories.

It is well known that clustering methods do not yield unique results, nor necessarily reproducible results for a stochastic method \cite{barbakh2009review}, and this indeterminacy will carry over path classification.
Nevertheless, any reasonable clustering can be expected to yield reasonable classes for mechanistic analysis and, moreover, \emph{any} classification will provide a ``fingerprint'' enabling quantitative comparison between different ensembles.



\subsubsection{Clustering and classification of discrete paths based on simple state sets} \label{sec:classVoro}
For future study, here we suggest a very simple distance metric based on state commonality for trajectories in discrete space.
In fact, this approach discards sequence information.
Before computing any distance between two trajectories, say, $\mathbf{s} = (s_1, s_2, ..., s_m)$ and $\mathbf{q} = (q_1, q_2, ..., q_n)$ where $s_1, q_1 \in A$ and $s_m, q_n \in B$, we are going to transform them to the sets $\mathcal{S} = \bigcup^m_{i=1} \{s_i\}$ and $\mathcal{Q} = \bigcup^n_{i=1} \{q_i\}$ respectively.

Now we define the distance $d$ between $\mathcal{S}$ and $\mathcal{Q}$ as 

\begin{equation}\label{eq:distance}
d(\mathcal{S},\mathcal{Q}) = \vert (\mathcal{S} \setminus \mathcal{Q}) \cup (\mathcal{Q} \setminus \mathcal{S}) \vert,
\end{equation}
where $|.|$ is the cardinality operator, $\setminus$ is the set difference (i.e., $A\setminus B = \{x: x \in A\; \mathrm{and} \;x \notin B \}$). In simple words, $d(\mathcal{S},\mathcal{Q})$ is the number of non-common elements between the sets $\mathcal{S}$ and $\mathcal{Q}$. 

The distance $d$ is also called symmetric difference between the sets $\mathcal{S}$ and $\mathcal{Q}$ and it is a true mathematical distance (i.e., $d(\mathcal{S},\mathcal{Q}) \geq 0$, $d(\mathcal{S},\mathcal{Q})=d(\mathcal{Q},\mathcal{S})$, $d(\mathcal{S},\mathcal{Q}) =0 \Longleftrightarrow  \mathcal{S} = \mathcal{Q}$ and $d(\mathcal{S},\mathcal{Q}) \geq d(\mathcal{S},\mathcal{R}) + d(\mathcal{R},\mathcal{Q})$)[ref]. However, what makes $d$ special is that it is a simple and very efficient metric to define Voronoi cells. 

In order to generate candidates for the Voronoi centers we first do hierarchical clustering based on $d$ and then the centroid $\zeta_k$ of the cluster $C_k$ is selected as shown in Eq.~\ref{eq:distance2}. The set $\{\zeta_k\}$, together with $d$ will completely determine to which class any path belongs to.

\begin{equation}\label{eq:distance2}
\zeta_k = \argmin_{\mathbf{s} \in C_k} \, \max_{\mathbf{q} \in C_k} d(\mathcal{S},\mathcal{Q}).
\end{equation}

In this case, paths are reduced to sets and, in contrast to the FS, the distance used is independent of the ordering of states in the path. The FS might be more intuitive and it is probably a better tool to interpret results. Also, discrete distances are not ideal for Voronoi-based classification since it is more likely to to find points equidistant to multiple Voronoi centers. Nevertheless, the clustering approach is equally valid for our goal of defining a partition of the path space that allows ``fingerprinting'' and comparison of different path ensembles.

\subsubsection{Clustering and classification of continuous trajectories}
As a final example for possible future work, we could use the Hausdorff or Fr\'{e}chet metric recently explored by Beckstein and coworkers \cite{Seyler2015} to classify an ensemble of continuous trajectories once the Voronoi centers in path space are defined. Although here we are recommending to use clustering as an intermediate step to define those centers, any other reasonable approach can be valid.

\subsection{Defining macrostates: Kinetic clustering of microstates based on commute times}
So far we have been talking about classification or clustering of pathways. 
However, implicit in pathways are the macrostates which transition pathways connect.
Here we cluster microstates in order to identify or define or macrostates, e.g., the folded and unfolded macrostates in a protein. There is no a unique way to do it, and that problem by itself could be the subject of another publication/research. Nevertheless, that is not the main goal of the present work and we are just going to propose one of the multiple ways in which the definition of macrostates can be done.

Here we employ a systematic, but fairly simple idea method building on prior work \cite{Chodera2007,zhang2010automated}.
The idea is to perform hierarchical kinetic clustering of MSM microstates, denoted as $i$ or $j$, while ensuring a separation of timescales.  
We can use as a distance metric the MFPT, or more precisely, the ``round-trip'' time, $t_{ij}$ = MFPT($i\to j$) + MFPT($j\to i$).  
A slight limitation of the implementation suggested below is the use of (inexact) Markovian MFPTs, because the unbiased direct estimation would be very noisy.

We employ a simple procedure: a hierarchical (or progressive) clustering based on a cutoff $t_{cut}$ and a factor $m$ defining the targeted separation of timescales. If the round-trip time ($t_{ij}$) is less that $t_{cut}$ then we merge the states.  Here is the proposed procedure:

\begin{enumerate}
\item Compute MFPT matrix $M$ and add it to $M^T$ to obtain the round-trip times $\{ t_{ij} \}$
\item While $\min(\{ t_{ij} \}) < t_{cut}$:
	\begin{itemize}
	\item Merge the corresponding states
	\item Recompute $\{ t_{ij} \}$ (step 1) for merged states
    
	\end{itemize}
\item Increase $t_{cut}$ until the maximum remaining element of $t_{ij}$  becomes less than a pre-defined multiple $m$ (e.g., 10) of $t_{cut}$. Use as macrostates the two clusters from the prior step with the maximum $t_{ij} > m \, t_{cut}$
\end{enumerate}
\section{Results}
\subsection{Classification based on the FS}
In this section we demonstrate the utility of the FS-based classification for the analysis and interpretation of complex stochastic pathways. We start with a two-dimensional toy model where it is easy to illustrate the how this method simplifies, in a meaningful way, the path ensemble. Then, we will follow the same steps on a much more complex system, protein B folding and unfolding from atomistic simulation.

\subsubsection{Toy Model}
\begin{figure}[h] 
\begin{center}
  \includegraphics[width=0.8\linewidth]{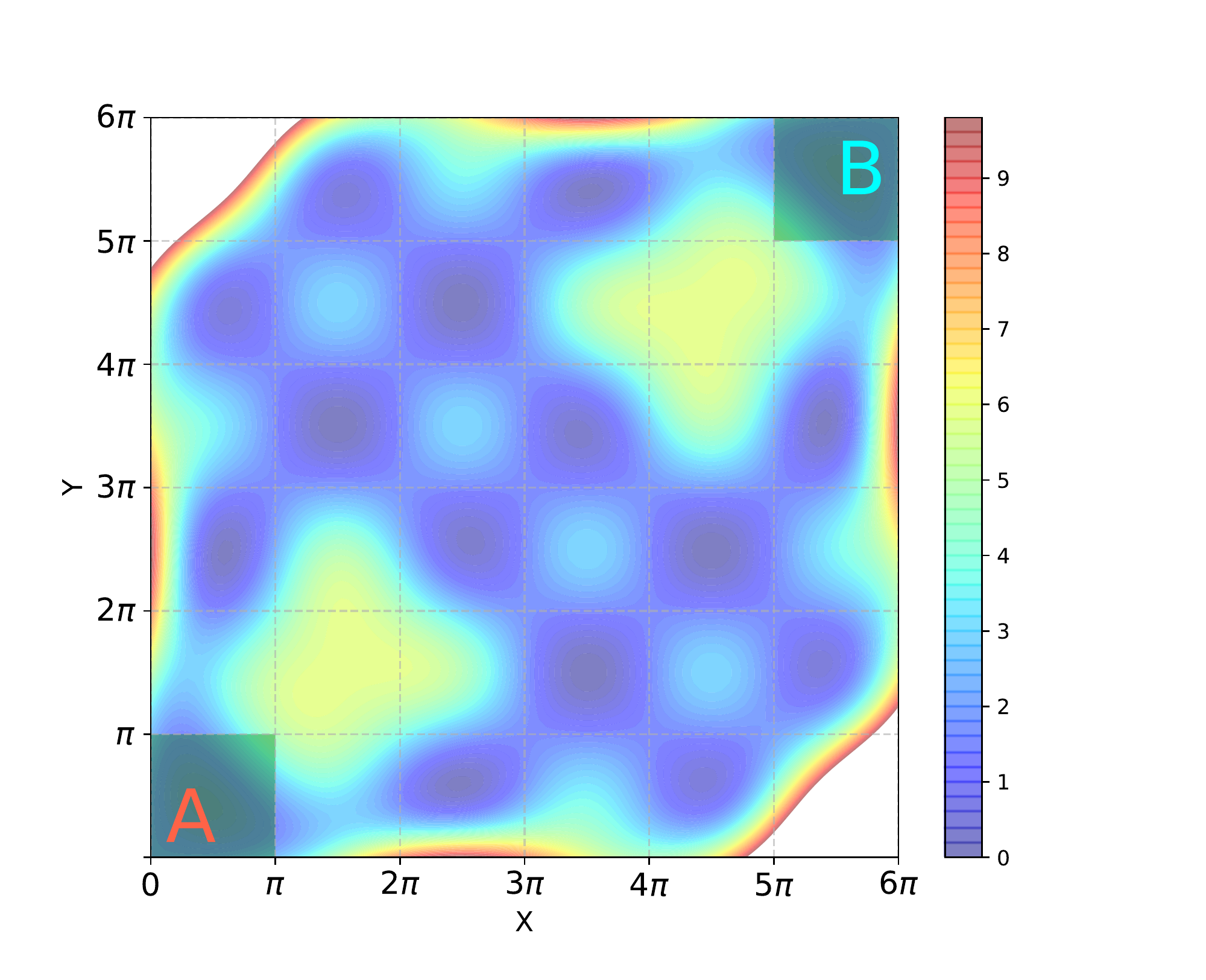}
  \caption{Simple 2D toy model. The energy values are in $k_BT$ units.}.
  \label{fig:ToyModel2D}
\end{center}
\end{figure}

Our toy model is a two-dimensional system with an energy function given by:

\begin{equation}
	E_{2D}/k_BT = \left\{ \begin{array}{lr}
 	\frac{3}{2}f(x,y) + \frac{9}{10^4}g(x,y)^2 g(y,x)^2 & \mbox{if}  \:\;0<x,y <6\pi \\
	\infty & \mbox{otherwise,} 
	\end{array}\right.
    \label{eq:toyEnergy}
\end{equation}
where $f(x,y)=1-\sin(x)\sin(y)$ and $g(x,y)=x -y - 9\sin(y/3)$. Fig.~\ref{fig:ToyModel2D} show the shape of the potential energy function $(E_{2D}/k_BT)$.

This is in some way a more complex version of our first example (see Fig.~\ref{fig:ToyModel}). Intuitively, we would have two dominant pathways or ``mechanisms''. However, in order to do the classification based on the FS, the model has to be discrete.

To analyze paths in this model, we built a (discrete) Markov model based on a (continuous-space) Monte Carlo simulation.
Simulations of around $10^6$ steps were done on the energy surface given in Eq.~\ref{eq:toyEnergy} using trial moves $(\delta x_i, \delta y_i)$ taken randomly from a uniform distribution in the region $[-\pi,\pi]\times [-\pi,\pi]$. The Markov states were defined by the squares $[k\pi,k\pi+\pi)\times [m\pi,m\pi+\pi)$, where $(k,m)$ is any combination of integers in $[0,5]$. For the path analysis we are also considering an initial state $A=\{(x,y) \in [0,\pi)^2\}$ and a final state $B=\{(x,y) \in [5\pi,6\pi)^2\}$ (see Fig.~\ref{fig:ToyModel2D}).

From now on, our analysis will focus on the discrete states of Markov model instead of the original continuous energy surface. For that reason the quality of the Markov model as a representation of the original model is not the main concern. Our goal for now is the classification of discrete paths and not how they where obtained. For visualization purpose, however, we are going to map discrete sequences back to the original energy surface.

\begin{figure}[h] 
\begin{center}
  \includegraphics[width=0.8\linewidth]{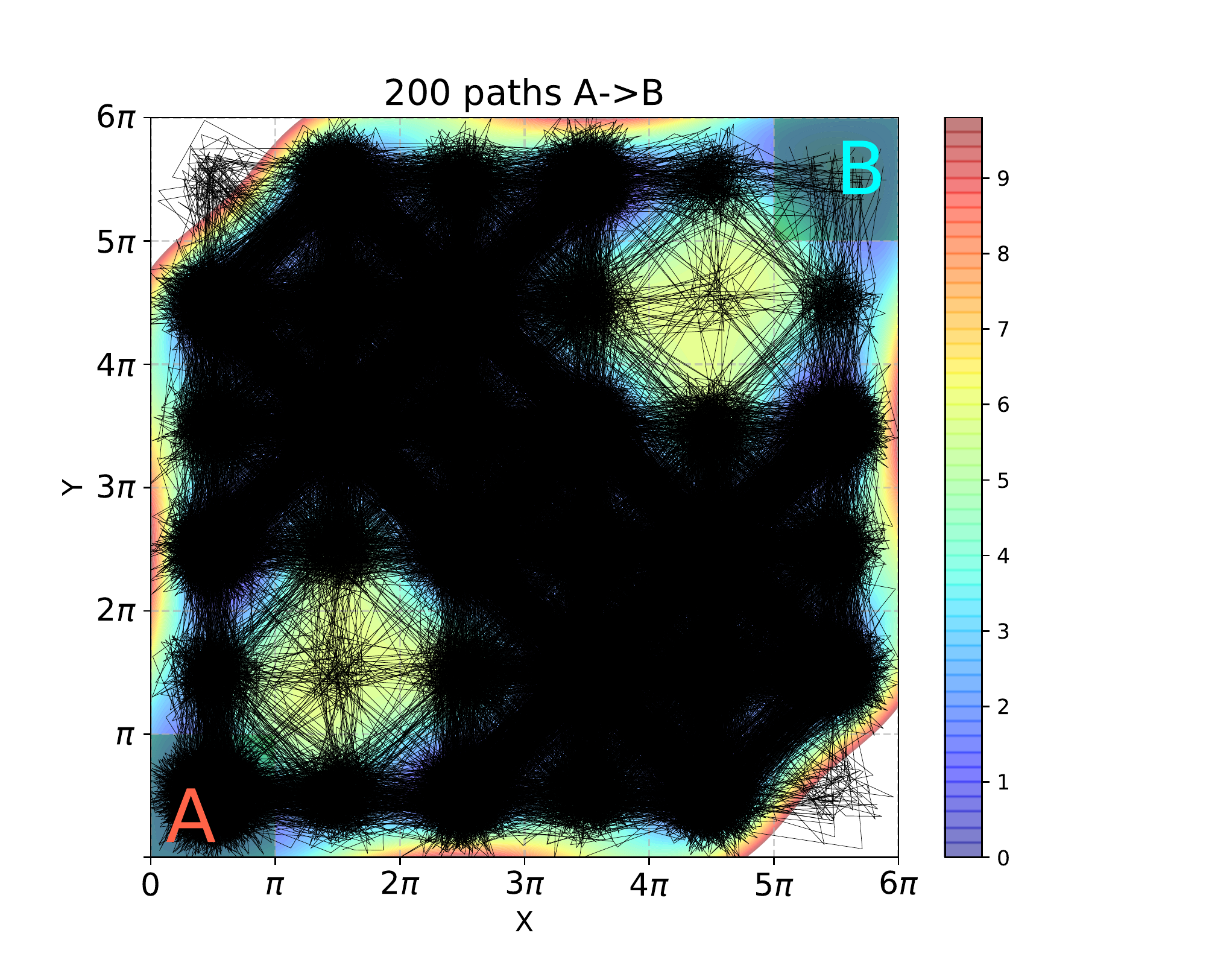}
  \caption{200 simulated paths from $A$ to $B$. The paths were generated from a discrete Markov model that \emph{learned} from the original trajectory and mapped back to the original energy surface. Noise were added just for visualization in order to avoid perfect overlap between discrete trajectories. The energy values are in $k_BT$ units.}.
  \label{fig:ToyModel2DPaths}
\end{center}
\end{figure}

Starting from the state $A$ (now a discrete index), $200$ discrete trajectories were generated until they reach $B$. The discrete trajectories are mapped, just for visualization, to the original energy surface, in particular to the center of each state ($[k\pi,k\pi+\pi)\times [m\pi,m\pi+\pi)$). The result, after adding noise in order avoid perfect overlap between distinct discrete trajectory fragments, is shown in Fig.~\ref{fig:ToyModel2DPaths}.  Due to the stochastic nature of the paths it is almost impossible to see some kind of ``structure'' in the path ensemble, even when the model was designed to have two dominant paths.

\begin{figure}[h] 
\begin{center}
  \includegraphics[width=0.8\linewidth]{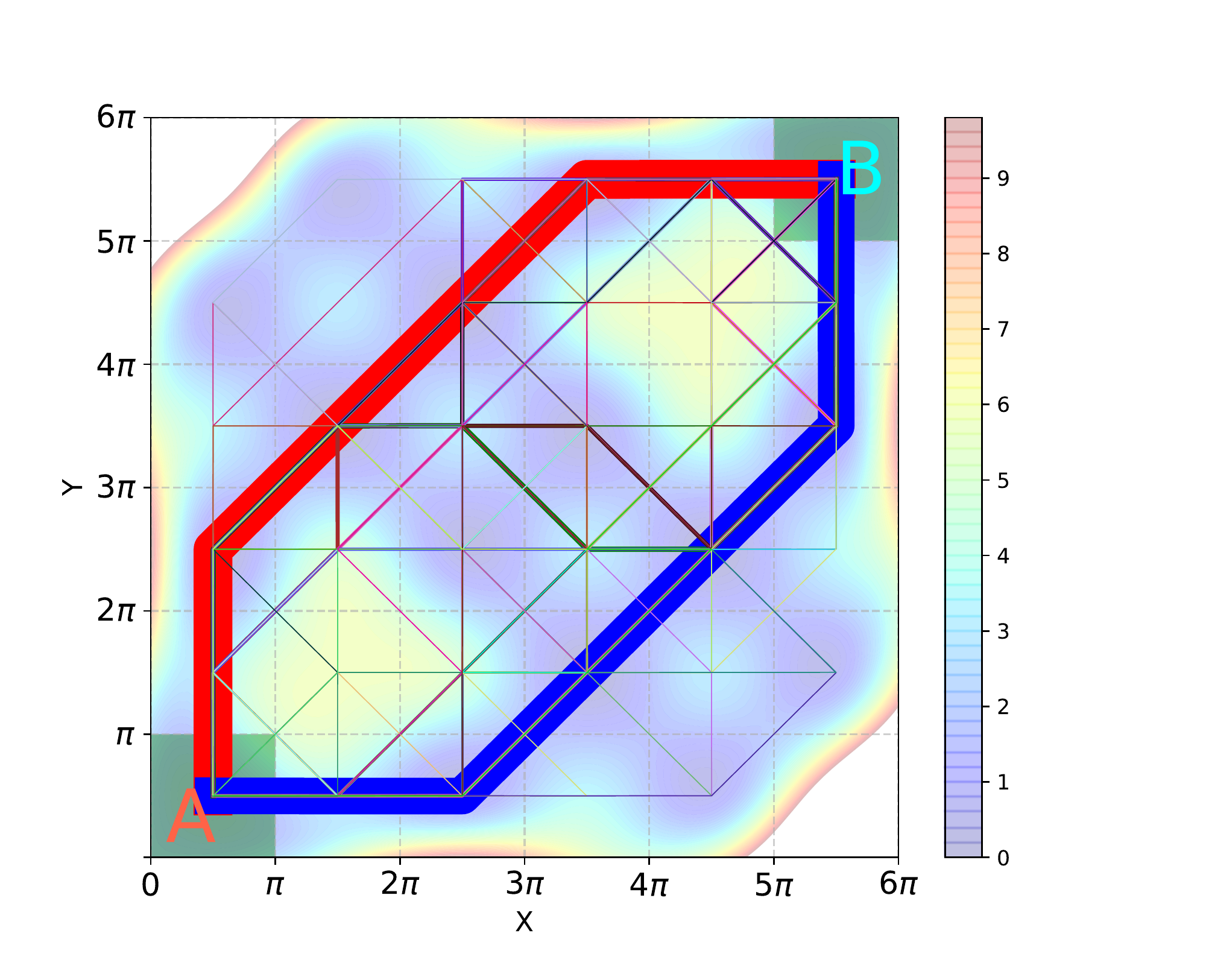}
  \caption{Path representation using fundamental sequences (FSs).  The 200 paths from $A$ to $B$ shown in Fig.\ \ref{fig:ToyModel2DPaths} are mapped to their fundamental sequences, plotted over the original energy surface.  The thickness of the FSs is proportional to their probabilities. The energy is in $k_BT$ units.}
  \label{fig:ToyModel2DFS}
\end{center}
\end{figure}

One the other hand, the classification based on the FS can clearly simplify the interpretation of the stochastic behavior of our model. Fig.~\ref{fig:ToyModel2DFS} shows the FSs plotted on the original energy surface; being the thickness of the individual FSs proportional to their observed probability. Notice that the FSs inherit the symmetry of the original model but we cannot expect a perfect symmetry since they are obtained by statistically.

\begin{figure}[h] 
\begin{center}
  \includegraphics[width=0.8\linewidth]{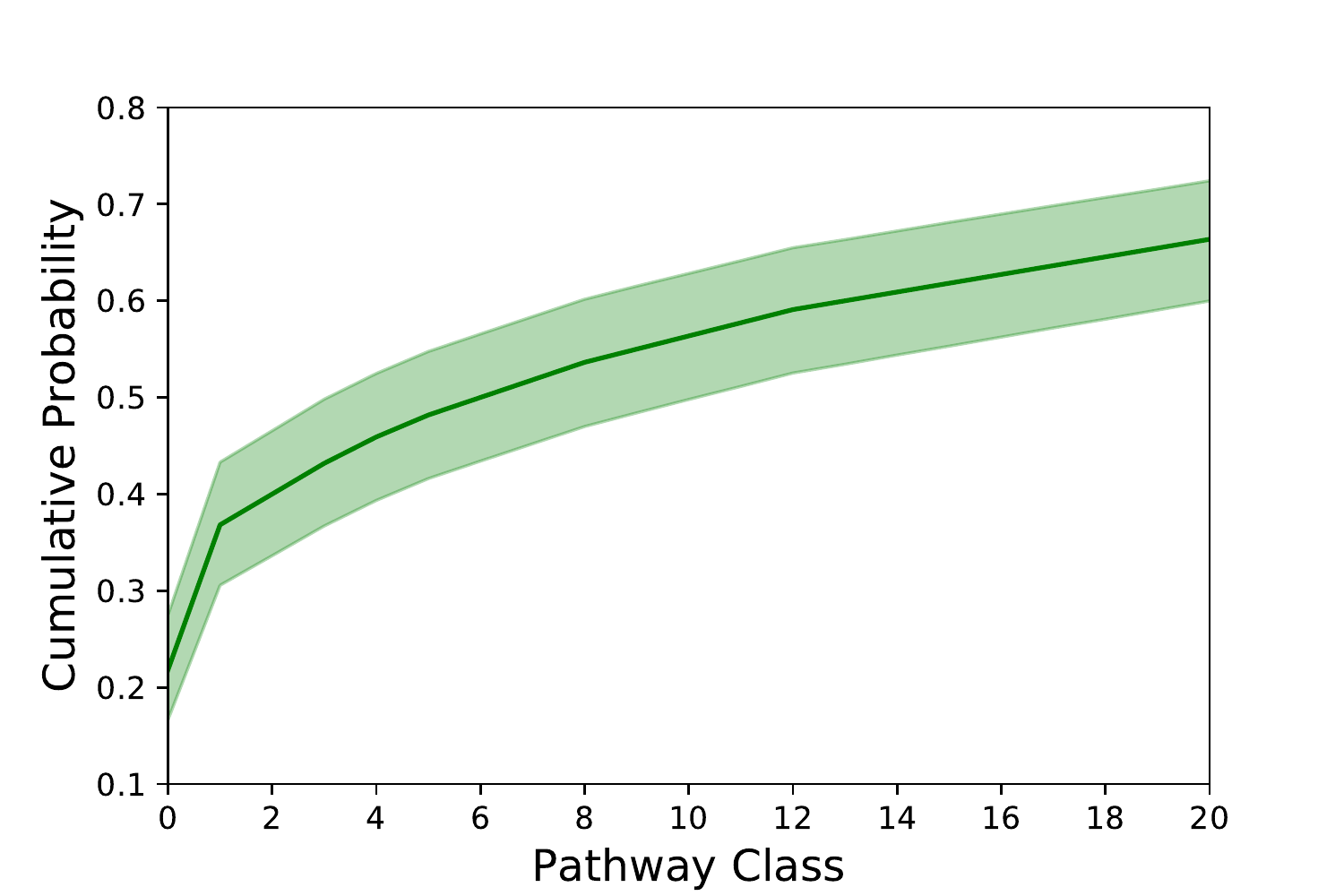}
  \caption{Toy-model pathway histogram.  The plot shows the summed histogram - i.e., the cumulative distribution of the classes defined by fundamental sequences, the light green region represents the 95\% confidence interval.  Data are derived from the 200 paths shown in Fig.\ \ref{fig:ToyModel2DPaths}.}
  \label{fig:CDF_ToyModel2D}
\end{center}
\end{figure}

The cumulative distribution of the FSs from $A$ to $B$ is shown in Fig.~\ref{fig:CDF_ToyModel2D}. The pathway classes are ordered by their preponderance, i.e., the lower the index the more probable. 
In this simple case, the two dominant FSs represent about one third of the path ensemble, and only the first $20$ pathway classes have more than one observed event. 
If there were another model of the system that we wanted to evaluate in terms of the path ensemble, we would just have to compare its distribution of the \emph{same} FSs, with those shown Fig.~\ref{fig:CDF_ToyModel2D}.

\subsubsection{Protein B atomistic (un)folding trajectories}

Next we study an atomistic, explicitly solvated model of Protein B, a much more complex system with $47$ residues that was simulated using regular molecular dynamics for $104\mu$s, enough time to observe multiple folding/unfolding events \cite{Lindorff-Larsen2011}.
As noted above, trajectories in a continuous space can be treated by a cluster-then-classify strategy applied in path space.
Here, however, we recapitulate the strategy used in the toy model of projecting configurations into a discrete space and using the fundamental-sequence analysis.

We started by reducing the configurational space of the system down to 40 discrete states using pyEMMA \cite{scherer2015pyemma}, yielding the space in which FSs will be computed.
We used the same states to build a non-Markovian (NM) model \cite{suarez2016accurate}, which can also be described as a ``history-augmented MSM.''
The folded and unfolded states where defined by a cutoff $\,t_{cut}=45$ns following the kinetic clustering approach described above.

\begin{figure}[h] 
\begin{center}
  \includegraphics[width=0.8\linewidth]{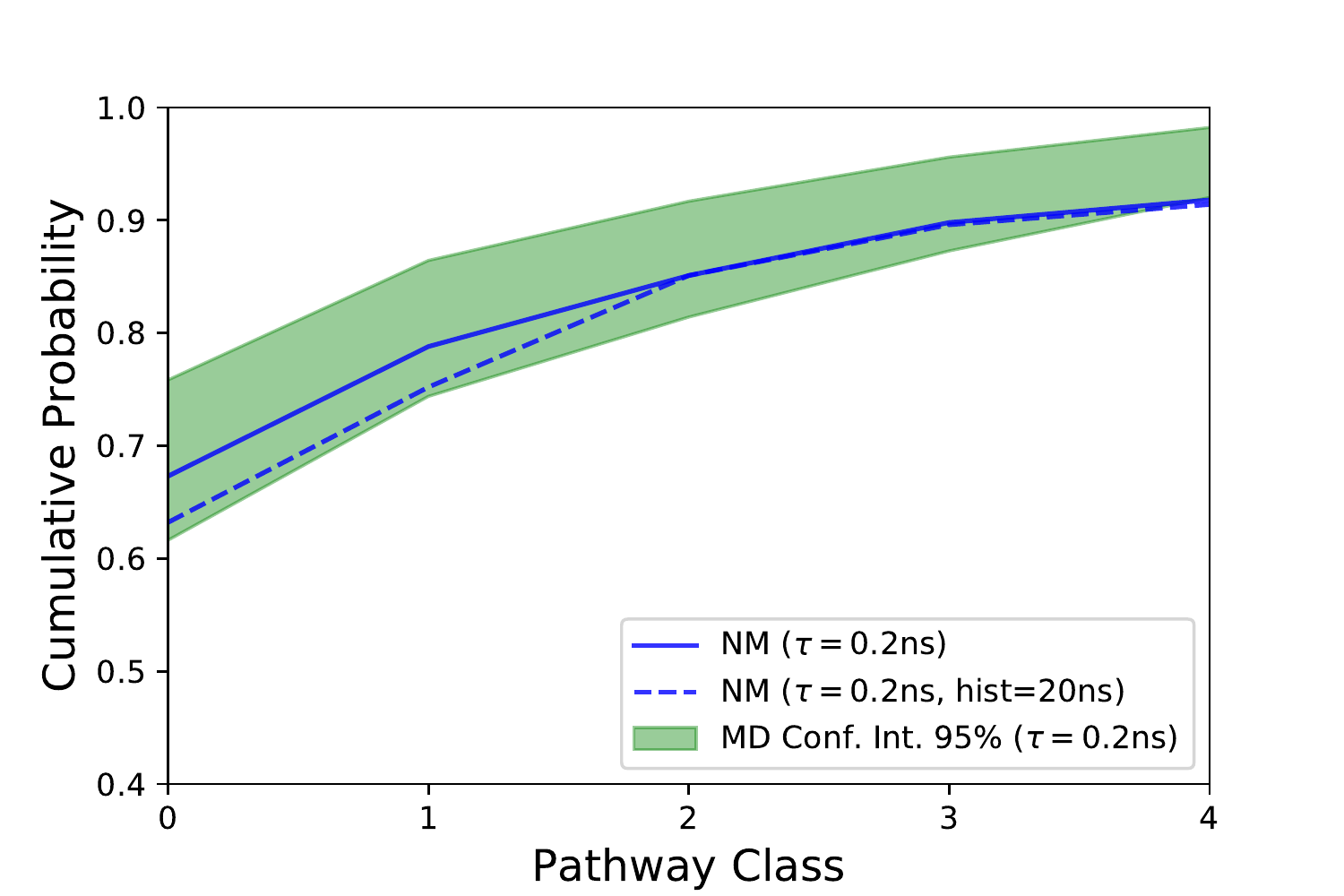}
  \caption{Protein B pathway histogram.  The cumulative distribution (summed histogram) of the fundamental-sequence classes is shown. The reference 95\% confidence interval of the class distribution based on MD data is shown as the green regaion.  Distributions obtained from the non-Markovian model with full history information (solid blue) and 20ns of history (dashed blue) are also shown.}
  \label{fig:CDF_PRB}
\end{center}
\end{figure}

To illustrate the value of the pathway histogram as a yardstick, we compare the results obtained from (a) the actual MD simulation analyzed using the discrete MSM states mentioned above, with two NM models, (b) one using all the history available and (c) another NM model built using only 20ns of history, as described previously\cite{suarez2016accurate}
See Fig.~\ref{fig:CDF_PRB}. 
As we can see, the NM models reproduce reasonably well the populations of at least the first five FS, representing almost $90\%$ of the path ensemble. 
The folding/unfolding mechanisms -- which are symmetric \cite{bhatt2011beyond} -- are dominated in this classification by a single FS (index 0) with almost 70\% of the statistical weight.

\begin{figure}[h] 
\begin{center}
  \includegraphics[width=1.1\linewidth]{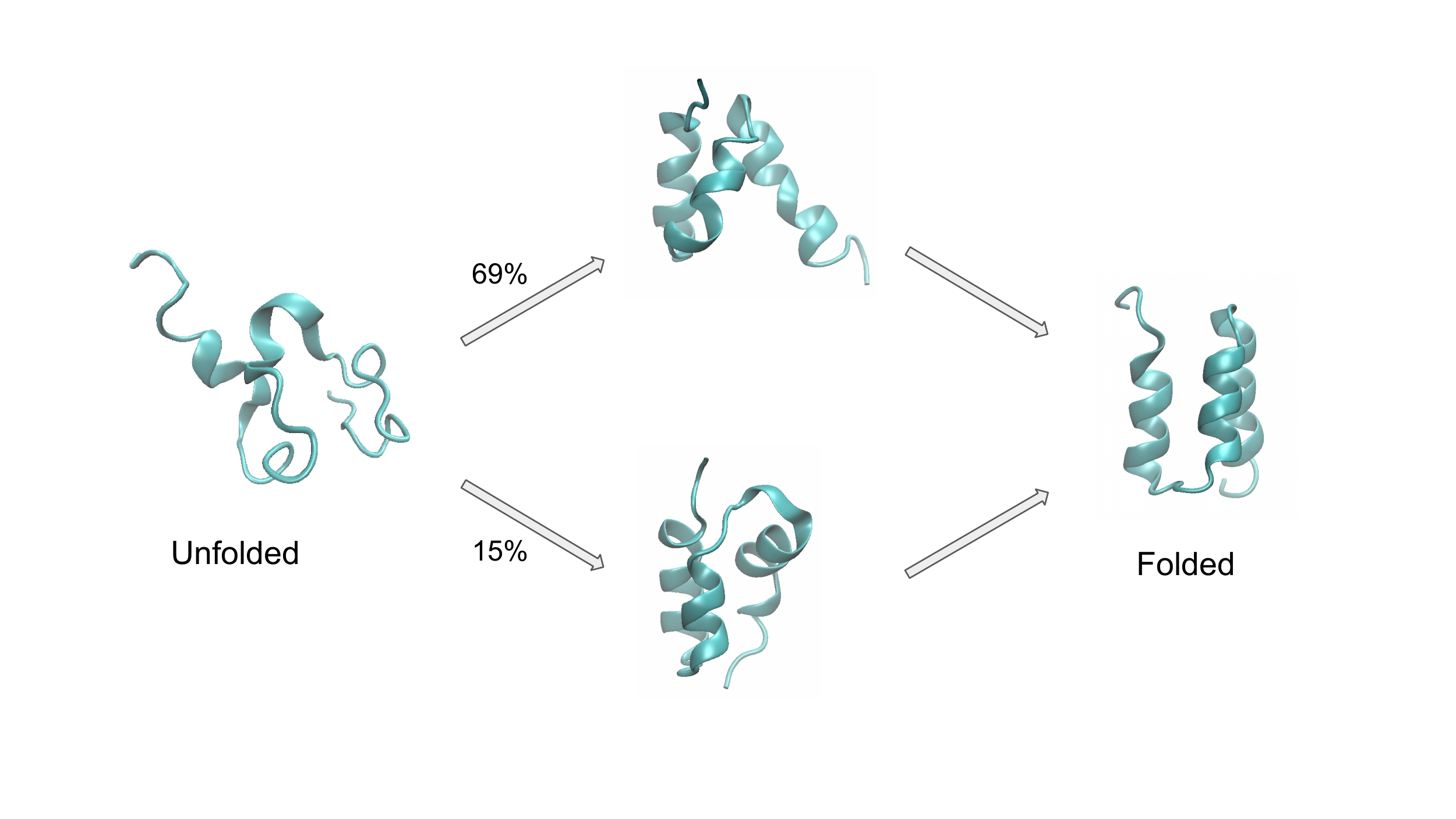}
  \caption{Primary Protein B folding pathways.  Representative intermediate structures are shown from the top two fundamental-sequence classes of Fig.\ \ref{fig:CDF_PRB}, along with the percent of transitions taking each pathway.}
  \label{fig:mechanism}
\end{center}
\end{figure}

Fig.~\ref{fig:mechanism} is a simplified representation of the two dominant pathways where each structure -- folded, unfolded and intermediate -- is a representative element of the corresponding pathway class. The intermediate structures suggest the FS analysis has automatically extracted truly distinct pathways.  We note that percentages in Fig.\ \ref{fig:mechanism} include data for folding only, and hence differ slightly (due to noise) from the data of Fig.\ \ref{fig:CDF_PRB}.

We emphasize that other classification schemes could be used, so our findings are not unique.
However, we expect the results to be structurally reasonable given the underlying basis in the clustered states derived from pyEMMMA \cite{scherer2015pyemma}.
Future work will investigate the degree to which path histograms can be considered ``robust'' -- i.e., insensitive to reasonable perturbations to the underlying featurization.
In any case, the FS analysis can bring quantitative insight to the understanding of complex dynamical processes like protein folding, as well as providing a standard yardstick for comparisons.


\clearpage
\section{Conclusions}
Although transition pathways have been analyzed in the past by a number of quantitative approaches \cite{Noe2009,Hummer2009,Seyler2015}, no universally applicable analysis has previously been introduced, to our knowledge.
Here we have provided employed a simple statistical structure, the histogram, along with a very general strategy for classifying trajectories into bins -- a strategy applicable to true stochastic trajectories whether in discrete or continuous space.
This histogram ``fingerprint,'' although not unique, enables a standardized comparison of mechanisms predicted by different models and/or simulation methods.

The pathway histogram can be constructed in arbitrary ways, and we have suggested several approaches for both discrete and continuous-space trajectories.
Notably, the automated ``fundamental sequence'' (FS) classification of discrete-state trajectories removes less informative pathway loops while retaining the potentially stochastic path ``backbone'' that is unlikely to follow flux lines.
Continuous-space trajectories can employ the FS classification scheme if configurations are first mapped to discrete states, as we have done.
Further, any clustering technique can be converted to a classification scheme using a Voronoi procedure; the only requirement is a distance metric in path space on which to base the clustering.

The pathway histogram approach is very generic and could equally be applied to other fields studying stochastic transitions in complex systems, including systems biology \cite{morelli2009dna,donovan2016unbiased,tse2018rare}.

\begin{acknowledgement}
We greatly appreciate constructive comments on this work from John Chodera and Jeremy Copperman.
We thank D.E. Shaw Research for providing access to the protein folding trajectory data set. This work was partially supported by NIH Grants R01GM115805 and P41GM103712, as well as NSF Grant MCB-1119091.

\end{acknowledgement}



\clearpage


\bibliography{pathway_histo_refs}
\end{document}